\documentclass[aps,prl,twocolumn,superscriptaddress,showpacs]{revtex4}
\usepackage{graphicx}
\usepackage{latexsym}
\usepackage{amssymb}
\usepackage{amsmath}
\usepackage{amsfonts}
\usepackage{bm}
\usepackage{multirow}
\usepackage{color}
\usepackage{slashed}
\usepackage{hyperref}
\usepackage{mathrsfs}
\usepackage{subfigure}
\usepackage{verbatim}
\usepackage{dsfont}
\usepackage{float}
\newcommand{\nc}{\newcommand}
\nc{\be}{\begin{equation}} \nc{\ee}{\end{equation}}
\nc{\bea}{\begin{eqnarray}} \nc{\eea}{\end{eqnarray}}
\nc{\bean}{\begin{eqnarray*}} \nc{\eean}{\end{eqnarray*}}
\nc{\dg}{\dagger} \nc{\ua}{\uparrow} \nc{\da}{\downarrow}

\newcommand{\vect}[1]{{\bm{#1}}}

\newcommand{\figref}[1]{Fig.\,\ref{#1}}

\newcommand{\beq}{\begin{equation}}
\newcommand{\eeq}{\end{equation}}
\newcommand{\beqn}{\begin{eqnarray}}
\newcommand{\eeqn}{\end{eqnarray}}

\begin{document}

\title{Bilayer Graphene as a Platform for Bosonic Symmetry Protected Topological States}

\author{Zhen Bi}
\affiliation{Department of physics, University of California,
Santa Barbara, CA 93106, USA}
\author{Ruixing Zhang}
\affiliation{Department of Physics, The Pennsylvania State
University, University Park, Pennsylvania 16802-6300, USA}
\author{Yi-Zhuang You}
\author{Andrea Young}
\affiliation{Department of physics, University of California,
Santa Barbara, CA 93106, USA}
\author{Leon Balents}
\affiliation{Kavli Institute for Theoretical Physics, University
of California, Santa Barbara, CA 93106-4030, USA}
\author{Chao-Xing Liu}
\affiliation{Department of Physics, The Pennsylvania State
University, University Park, Pennsylvania 16802-6300, USA}
\author{Cenke Xu}
\affiliation{Department of physics, University of California,
Santa Barbara, CA 93106, USA}

\date{\today}

\begin{abstract}

Bosonic symmetry protected topological (BSPT) states, the bosonic
analogue of topological insulators, have attracted enormous
theoretical interest in the last few years. Although BSPT states
have been classified by various approaches, there is so far no
successful experimental realization of any BSPT state in two or
higher dimensions. In this paper, we propose that a two
dimensional BSPT state with $U(1) \times U(1)$ symmetry can be
realized in bilayer graphene in a magnetic field. Here the two
$U(1)$ symmetries represent total spin $S^z$ and total charge
conservation respectively. The Coulomb interaction plays a central
role in this proposal -- it gaps out all the fermions at the
boundary, so that only bosonic charge and spin degrees of freedom
are gapless and protected at the edge. Based on the bosonic nature
of the boundary states, we derive the bulk wave function for the
bosonic charge and spin degrees of freedom, which takes exactly
the same form as the desired BSPT state. We also propose that the
bulk quantum phase transition between the BSPT and trivial phase,
could become a ``bosonic phase transition" with interactions. That
is, only bosonic modes close their gap at the transition, which is
fundamentally different from all the well-known topological
insulator to trivial insulator transitions which occur for free
fermion systems. We discuss various experimental consequences of
this proposal.

\end{abstract}


\maketitle

A symmetry protected topological (SPT) state, first defined in
Ref.~\onlinecite{wenspt,wenspt2}, is the ground state of a local
quantum many-body Hamiltonian whose bulk is gapped and
nondegenerate, but whose boundary remains either gapless or
degenerate as long as the entire system including the boundary
preserves certain symmetries. Fermionic SPT states include the
familiar quantum spin Hall (QSH)
insulator~\cite{kane2005a,kane2005b}, the three-dimensional ($3d$)
topological insulator (TI)~\cite{fukane,moorebalents2007,roy2007},
and topological superconductors. Noninteracting fermionic SPT
states have been fully classified and
understood~\cite{ludwigclass1,kitaevclass}. Unlike fermionic
systems, bosonic SPT (BSPT) states require strong interaction to
overcome the tendency to form Bose-Einstein condensates. The
simplest and most well-known BSPT state is the $1d$ Haldane phase,
which can be realized in the simplest nearest-neighbor spin-1
Heisenberg chain~\cite{haldane1,haldane2}. However, higher
dimensional generalizations of BSPT states have not been found.
The only even {\em potentially feasible} experimental proposal is
for a bosonic integer quantum Hall state in ultracold
atoms~\cite{senthillevin}, but even this seems far away, since as
yet experiments with both rotating traps and artificial magnetic
fields are still far from the quantum Hall regime. The exactly
soluble parent Hamiltonians constructed in
Ref.~\onlinecite{wenspt,wenspt2} in dimensions higher than one all
involve high order multiple spin interactions, and are thus
unlikely to exist in realistic materials. Up to now, all
approaches to classifying and characterizing BSPT
states~\cite{wenspt,wenspt2,senthilashvin,luashvin,xuclass,wangguwen}
rely on mathematical or effective field theory descriptions, which
shed little light on how to identify a realistic candidate BSPT
state.

In the current paper, we hope to bridge the gap between
theoretical studies and experimental realizations of BSPT states.
We propose that bilayer graphene in magnetic field (with both
inplane and out-of-plane components) provides a platform of
realizing and probing the $2d$ BSPT state with $U(1)_s \times
U(1)_c$ symmetry, where $U(1)_s$ and $U(1)_c$ correspond to the
total spin$-S^z$ and total electric charge conservation
respectively. Based on the formalism developed in
Ref.~\onlinecite{luashvin,xuclass}, this state has a $\mathbb{Z}$
classification, $i.e.$ with these symmetries there is an infinite
set of non-trivial $2d$ BSPT classes, which are indexed by an
integer $k$.
Effective field theory descriptions of these BSPT states have been
given in terms of
Chern-Simon field theory~\cite{luashvin} and a non-linear sigma
model (NLSM) with a $\Theta$-term~\cite{xusenthil,xuclass}.
The action for the latter is
\begin{equation}
\mathcal{S} = \int d^2x d\tau \ \frac{1}{g} (\partial_\mu
\vect{n})^2 + \frac{i \Theta}{\Omega_3} \epsilon_{abcd} n^a
\partial_x n^b \partial_y n^c \partial_\tau n^d, \label{o4}
\end{equation}
where $\vect{n} = (n_1, n_2, n_3, n_4)$ is a four component vector
with unit length~\cite{xusenthil,xuclass}, and $\Omega_3$ is the
volume of a $3d$ sphere with unit radius. In Eq.~\ref{o4}, the
BSPT phases correspond to the strongly interacting fixed point $g
\rightarrow \infty$, and $\Theta \rightarrow 2k\pi$ with nonzero
integer $k$, while the trivial phase corresponds to the fixed
point $\Theta \rightarrow 0$. The quantum phase transition between
different BSPT phases is driven by tuning $\Theta$ in
Eq.~\ref{o4}, and the critical point is at $\Theta = (2k+1)\pi$. A
similar phase diagram and renormalization group flow for NLSMs in
one lower dimension was studied thoroughly in
Ref.~\onlinecite{pruisken1,pruisken2}.

\begin{figure}[tbp]
\begin{center}
\includegraphics[width=240pt]{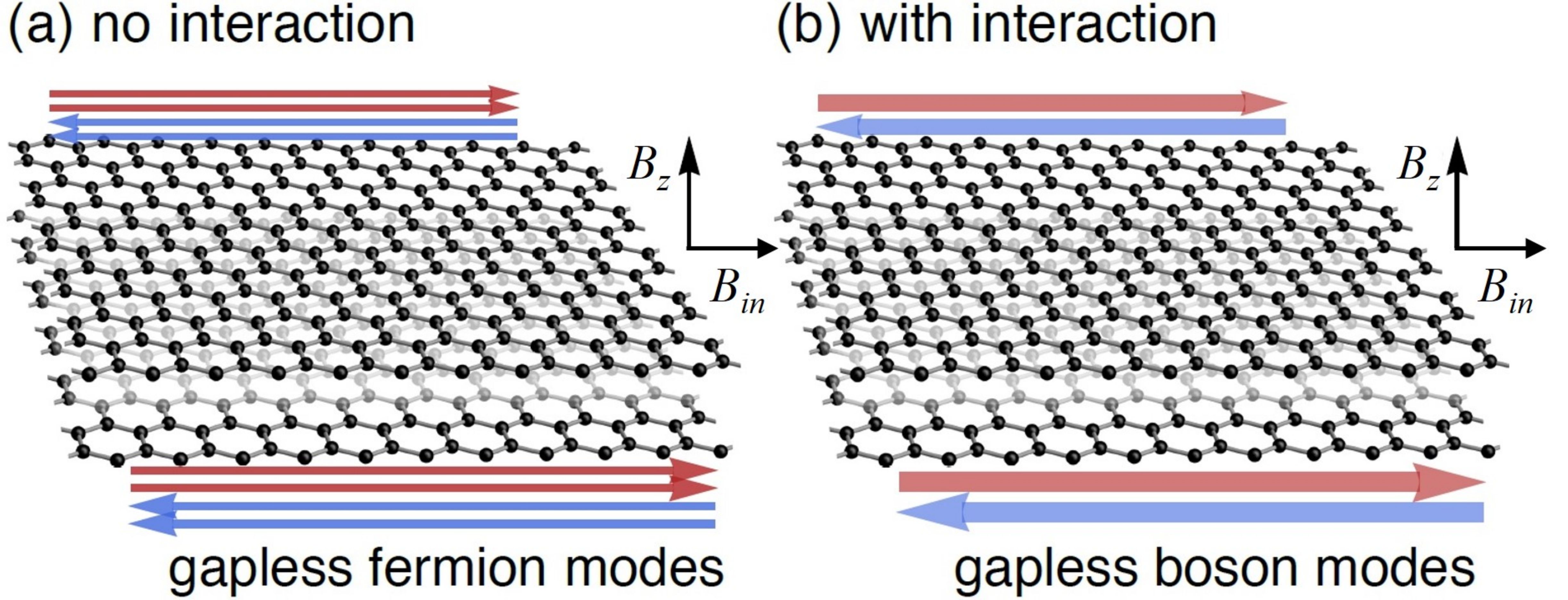}
\caption{Schematic of bilayer graphene in the presence of a
magnetic field with both inplane and out-of-plane components. (a)
Without interactions, the boundary hosts two channels of fermionic
edge states with total central charge $c = 2$. (b) Including the
Coulomb interactions, there is only one gapless channel of bosonic
edge state with $c = 1$.} \label{fig: graphene}
\end{center}
\end{figure}

Let us elaborate on our claim. It was proposed that an
out-of-plane magnetic field drives undoped graphene into a
``quantum spin Hall insulator"\cite{abaninlee} (it is also called
the ferromagnetic quantum Hall state, since the bulk is fully spin
polarized. In order to avoid a canted antiferromagnetic phase, one
also needs an inplane magnetic field to increase the Zeeman
coupling~\cite{caftheory,young2013}, which will be discussed in
detail in the supplementary material~\footnote{Please see the
supplementary material, which includes
Ref.~\onlinecite{maxim,Gonz,young2016,martin2005,maciejko2009,kane1992}.}).
In a bilayer, this possesses at the Hartree-Fock level two
channels of counter-propagating spin-filtered helical fermionic
edge states~\cite{edgec,young2013}. However, when the Coulomb
interaction is included, we will demonstrate that (as illustrated
in \figref{fig: graphene}), the behavior is qualitatively modified
to correspond precisely to that of the BSPT theories,
Eq.~(\ref{o4}) with $k=1$, so that, although it is built from
electrons, it is a proper BSPT state in the following senses:

{\it 1.} the Coulomb interaction, which is expected to play an
important role in this system, induces a gap for all fermionic
excitations at the boundary, while bosonic charge and spin
excitations remain gapless and protected by the two $U(1)$
symmetries (Fig.~\ref{fig: graphene}$b$);

{\it 2.} after the fermions are gapped out at the boundary by the
Coulomb interaction, using the correlation functions of the
boundary states, and following the procedure in
Ref.~\onlinecite{cftblock}, one can derive the bulk wave function
for the bosonic charge and spin, which takes exactly the form as
the BSPT wave function constructed using the mutual flux
attachment picture in Ref.~\onlinecite{levinsenthil}.


{\it 3.} Using the Chalker-Coddington picture, the bulk quantum
phase transition between the nontrivial SPT phase ($k=1$) and
trivial ($k=0$) phase (hereafter phrased as ``topological to
trivial transition") can be described by percolation of domains
and the corresponding network of interface/boundary states.
Because the boundary only has gapless bosonic modes, such a
topological quantum phase transition can occur while preserving
the bulk gap for fermionic quasiparticles. The topological to
trivial transition can be driven by varying competing magnetic and
electric fields, and we propose that the bosonic scenario for this
quantum phase transition could occur with sufficiently strong
interactions. This is a qualitatively different situation from the
well-known topological to trivial transitions in weakly correlated
systems, such as the plateau transition between integer quantum
Hall states, or the transition between normal and topological band
insulators -- these transitions have a free fermion description
which involves the fermion gap closing in the bulk. The above
statement is supported by recent numerical studies of a similar
model on the bilayer honeycomb lattice~\cite{kevinQSH,mengQSH2}.



We now proceed to an exposition of these results. In this work we
will focus on the boundary states and the bulk wave function of
the BSPT state, we will defer the detailed analysis of the bulk
topological transition to future study. For non-interacting
bilayer graphene, there are two channels of helical edge states,
described by the Hamiltonian
\begin{equation} H_0 = \int dx \ \sum_{l=1}^2
\psi_{l,L}^\dagger i v \partial_x \psi_{l,L} - \psi^\dagger_{l,R}
i v
\partial_x \psi_{l,R},
\end{equation}
where $l = 1,2$ labels the channels, $L,R$ denote
the left and right moving fermions respectively, which also correspond
to electrons with spin-up and down, and $v$ is the Fermi velocity~\footnote{In
  principle the velocity of the two channels of edge states could be
  different, but this velocity difference would be unimportant for the
  rest of the analysis.}.  
The presence of {\em some} counter-propagating edge states was
deduced experimentally from non-local transport
signatures~\cite{young2013}. When the Coulomb interaction is
ignored, the boundary is a free fermion conformal field theory
(CFT) with central charge $c = 2$.


The free fermion edge states can be bosonized
into two flavors of free bosons:
\begin{equation}
 H_{0} = \int dx \
\sum_{l = 1}^2 \frac{v}{2K} (\partial_x \theta_l)^2 + \frac{v
K}{2} (\partial_x \phi_l)^2, \label{H0}
\end{equation}
where $[\theta_{l}(x), \
\partial_{x^\prime} \phi_{l^\prime}(x^\prime)] = i
\delta(x-x^\prime) \delta_{ll^\prime}$, and $\psi_{l, L/R} \sim
e^{i\theta_l \pm i \pi \phi_l}$. For free $1d$ fermions without
interaction, the Luttinger parameter $K = \pi$.

Coulomb interactions $H_{int}$ are conveniently handled in the
bosonization framework.  Using the representation of the fermion
density $n_l \sim \partial_x \phi_l$, one obtains:
\begin{equation}
H_{int} = \int dx \ \sum_{l=1}^2
\frac{U_{\rm intra}}{2} (\partial_x \phi_l)^2 + U_{\rm inter}
\partial_x \phi_1 \partial_x \phi_2 + H_{v}, \label{Hi}
\end{equation}
where
$U_{\rm intra}$ and $U_{\rm inter}$ represent intralayer and
interlayer forward-scattering interactions, respectively.  $H_v$ is an
anharmonic vertex term, and will play a central role
here~\footnote{Interaction can induce another anharmonic term:
$\cos(2\theta_1 - 2\theta_2)$, but this term is irrelevant in our
system.}:
\begin{equation}
H_v \sim \alpha \cos (2\pi \phi_1 - 2\pi
\phi_2).
\end{equation}
Here we have assumed that the long range Coulomb interaction is
screened to a short range one, but this is not essential.
Physically $H_v$ describes the backscattering between two channels
of edge states: $H_v \sim \psi^\dagger_{1,L}
\psi^{\vphantom\dagger}_{1,R} \psi^\dagger_{2,R}
\psi^{\vphantom\dagger}_{2,L}$. The anharmonic $H_v$ is relevant
in the renormalization group sense, as long as $U_{\rm intra} >
U_{\rm inter}$. This condition is naturally satisfied because
$U_{\rm inter}$ is suppressed by the square of the wave function
overlap between the two channels of edge states.


When it is relevant, $H_v$ will ``pin'' the bosonic mode
$\phi_- = (\phi_1 - \phi_2)/2$, causing large fluctuations of
$\theta_{-} = \theta_1 - \theta_2$, leading to a gap in this
antisymmetric sector, and also a gap for all fermions at the
boundary.  The symmetric edge modes
$\phi=(\phi_1 + \phi_2)/2$ and $\theta=\theta_1 + \theta_2$, however,
remain gapless, because $\theta$ transforms under symmetry $U(1)_c$,
while $\phi$ transforms under $U(1)_s$.  It is straightforward to show
-- see below -- that only physical operators which create bosonic
excitations can be built from the gapless $\phi,\theta$ fields,
consistent with the statement that the boundary has
symmetry protected gapless bosonic modes. 
The size of the fermion gap at the boundary state is estimated in
detail in the supplementary material.

The effective low energy theory that describes the canonical
conjugate modes $\phi$ and $\theta$ reads
\begin{equation}
\tilde{H} = \int dx \ \frac{
\tilde{v} }{2 \tilde{K}} (\partial_x \theta)^2 + \frac{\tilde{v}
\tilde{K}}{2} (\partial_x \phi)^2. \label{cft1}
\end{equation}
Hence interaction reduces the central charge of the system from
$c=2$ to $c=1$. Because $\theta$ and $\phi$ transform nontrivially
(i.e. shift under $U(1)_c$ and $U(1)_s$ symmetries respectively),
there are no anharmonic vertex operators allowed by symmetry in
Eq.~\ref{cft1}.  Because $\theta$ and $\phi$ are ``dual" to each
other, a unit soliton of $\phi$ at the $1d$ boundary carries
charge-$2e$, and a unit soliton of $\theta$ carries spin $S^z =
1$. The gaplessness of the boundary state is protected by the
$U(1)_c \times U(1)_s$ symmetry alone: even if the translation
symmetry of the boundary is broken by disorder (which is
inevitable in any real system), as long as the $U(1)_c \times
U(1)_s$ symmetry is preserved, the boundary must still remain
gapless. The edge state in our system is also very different from
the cases studied in Ref.~\onlinecite{valley1,valley2}, since in
those systems the states localized at the domain wall is unstable
to disorder.


Here we note that although the bosonization of the edge states of
bilayer graphene in a magnetic field was also studied in
Ref.~\onlinecite{fertig1,fertig2}, in these works only the spin
symmetry was considered in the bosonization, and the conclusion of
Ref.~\onlinecite{fertig1,fertig2} was that the system is equivalent to
a $1d$ spin model. Here we stress that, both the $U(1)_s$ and $U(1)_c$
symmetries are crucial to define the BSPT state: $i.e.$ if either of
the U(1) symmetries is broken (for example if the bulk forms a canted
antiferromagnetic order), the system will become a trivial state. With
both $U(1)$ symmetries in our system, the boundary theory
Eq.~\ref{cft1} must remain gapless, and it can never be realized as a
$1d$ system, but rather only as the boundary of a $2d$ system, which
is an essential property of all SPT states.

Let us discuss the operator content further. Assuming $\phi_-$ is
pinned and $\theta_-$ fluctuates strongly, one can obtain the low
energy components of the four component vector $\vect{n}$ in
Eq.~\ref{o4}:
\begin{eqnarray}
n_1 + in_2 &&  \sim
\epsilon_{\alpha\beta} \psi_{1, \alpha} \psi_{2,\beta}   \qquad \sim   e^{i
\theta},\nonumber \\
  n_3 + in_4 && \sim  \sum_l (-1)^l \psi^\dagger_l \sigma^+
\psi^{\vphantom\dagger}_l  \sim  e^{i 2\pi\phi}. \label{bf}
\end{eqnarray}
Here $n_1 + in_2$ corresponds to an interlayer spin-singlet
($S^z=0$) Cooper pair, while $n_3$ and $n_4$ correspond to
in-plane magnetic order.  All components of the vector $\vect{n}$
have power-law correlations at the boundary, and their scaling
dimensions are $\Delta[\epsilon_{\alpha\beta} \psi_{1, \alpha}
  \psi_{2,\beta}] = \frac{\tilde{K}}{4\pi}$, $\qquad \Delta[\sum_l (-1)^l
  \psi^\dagger_l \sigma^+ \psi^{\vphantom\dagger}_l] =
  \frac{\pi}{\tilde{K}}.$
Thus we see that indeed the low energy correlations at the edge
all correspond to bosonic fields, which could be built from
elementary bosons of even charge and integer spin. The presence of
four distinct ``primary fields'' is characteristic of the
Wess-Zumino-Witten (WZW) SU(2)$_1$ CFT, which is well-known to be
expressable in terms of a single gapless boson and has
$c=1$\cite{Witten1984,KnizhnikZamolodchikov1984}. The model in
Eq.~\eqref{cft1} is a deformation of the usual SU(2)$_1$ theory
which reduces the symmetry to U(1)$_c\times$U(1)$_s$. It is also
equivalent to a (deformed) O(4) NLSM with $k=1$ WZW term -- see
e.g. Ref.~\onlinecite{spn}.

Eq.~\ref{bf} identified the effective bosonic degrees of freedom
that form a bosonic SPT state in the bulk. There are two flavors
of bosons carrying charge and spin quantum numbers respectively.
Following the method of Ref.~\cite{cftblock}, we can derive the
wave function of the bosons in the bulk, by calculating the
following correlation function of the boundary conformal field
theory: \beqn \Psi(z_1, z_2 \cdots w_1, w_2 \cdots) \sim \langle
\prod_j e^{i\theta(z_j)} \prod_k e^{2\pi i \phi(w_k)}
\mathcal{O}_{bg} \rangle, \label{wf} \eeqn where $z_j$ and $w_k$
are the complex coordinates in the $2d$ plane for the two flavors
of bosons. This is equivalent to calculating the partition
function of a $2d$ Coulomb gas with both electric and magnetic
charges~\cite{kadanoff,francesco}, and $\mathcal{O}_{bg}$
represents a neutralizing background charge operator. The
correlation function in Eq.~\ref{wf} can be evaluated with either
Eq.~\ref{H0} or Eq.~\ref{cft1}, and the result will be
qualitatively the same: \beqn \Psi(z_1, z_2 \cdots w_1, w_2
\cdots) \sim \mathrm{Norm}(z_j, w_k) \prod_{j,k} (z_j - w_k),
\eeqn where $\mathrm{Norm}(z_j, w_k)$ only depends on the norm of
$z_j - w_k$, $z_i - z_j$ and $w_i - w_j$, and contains all the
dependence upon the Luttinger parameters in Eq.~\ref{H0} and
Eq.~\ref{cft1}. This wave function indeed represents a bosonic SPT
state: it is symmetric under interchange of identical $z_i$ or
$w_j$ bosons, and the two flavors of bosons view each other as a
$2\pi-$flux. This mutual ``flux attachment" picture is the very
essence of the BSPT state~\cite{levinsenthil}.

Knowing the effective field theory at the boundary is the $(1+1)d$
NLSM for $\vect{n}$ with a Wess-Zumino-Witten term at level $k=1$,
the bulk theory can be constructed with the Chalker-Coddington
network model~\cite{cc}, and as was shown in
Ref.~\onlinecite{senthilashvin,xuludwig}, the bulk theory obtained
by this construction is precisely Eq.~\ref{o4} with $\Theta =
2\pi$. The physical meaning of this topological $\Theta-$term is
that, a vortex of $(n_1, n_2)$, $i.e.$ a vortex of the
superconductor order parameter, which traps magnetic flux
$\frac{hc}{2e}$, would carry spin $S^z = 1$, which is perfectly
consistent with the physics of the bilayer QSH state.

It is worth contrasting with the case of a single layer QSH
insulator, in which the boundary {\it cannot} be driven into a
state with gapped fermions but gapless bosonic modes, as long as
the $U(1)_c$ and time-reversal (or $U(1)_s$) symmetry of the QSH
insulator are preserved~\cite{xuedge,wuedge}. The mapping between
fermionic QSH insulator and BSPT is only valid for two copies of
QSH insulators (which mathematically is equivalent to four copies
of $p \pm ip$ topological superconductors), as was shown in
Ref.~\onlinecite{xufb}.

By varying competing electric and magnetic fields normal to the
layer, a quantum phase transition can occur between the BSPT and
the trivial state in the $2d$ bulk. Using the Chalker-Coddington
network picture, one may construct a theory for the 2d bulk phase
transition which involves only gapless bosonic modes and retains
the single-fermion gap. In the field theory Eq.~\ref{o4} this
transition occurs when $\Theta$ is tuned to $\pi$. Although
directly analyzing the bulk field theory at $\Theta = \pi$ is
difficult, recent unbiased determinant quantum Monte Carlo
simulation on a similar bilayer honeycomb lattice interacting
fermion model confirms that this purely bosonic
topological-trivial quantum phase transition can indeed
happen~\cite{kevinQSH,mengQSH2}, which is fundamentally different
from the ordinary topological to trivial transition in any free
fermion system. Maintaining the single particle gap requires
strong interactions, and other less interesting possibilities are
possible in experiment, such as other intermediate phases between
the BSPT phase and the trivial phase. Nevertheless, a direct
second order ``bosonic" transition like the one found in
Ref.~\onlinecite{kevinQSH,mengQSH2} seems allowed and a quite
interesting prospect.

{\it Experimental Implications}

\begin{figure}[b]
\begin{center}
\includegraphics[width=250pt]{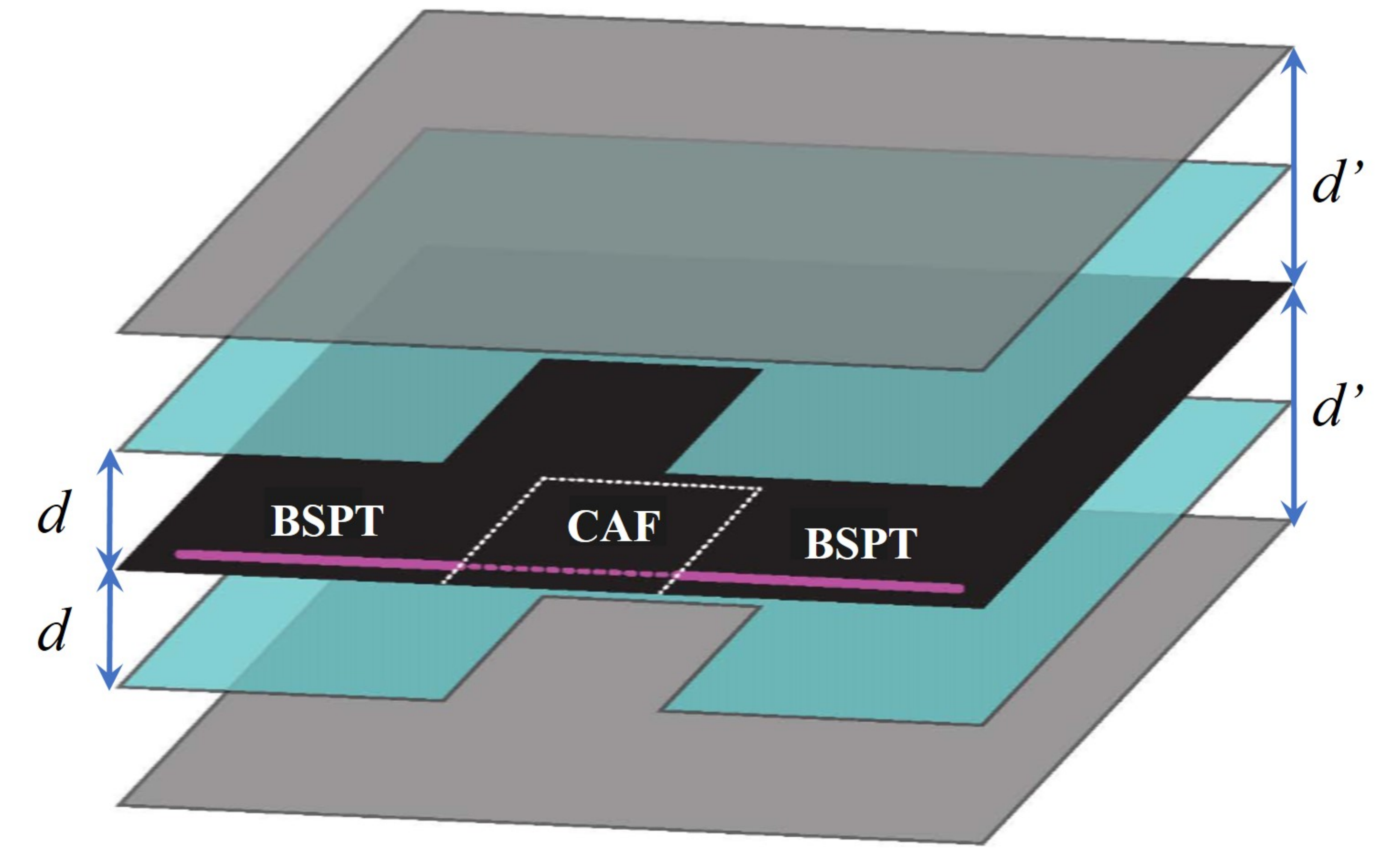}
\caption{Our proposed set-up for measuring the carrier charge at
the boundary of our system. Most of the sample are screened by the
inner symmetric gates, while the unscreened region has a stronger
interaction which leads to a CAF order, and induces backscattering
of the edge states. We also add a pair of outer gates to control
the strength of interaction in the CAF region.} \label{structure}
\end{center}
\end{figure}

The central prediction of our theory is that in a bilayer graphene
in the quantum spin Hall phase~\cite{young2013}, the gapless
boundary modes are bosonic rather than fermionic. The low energy
charge carriers on the edge are Cooper pairs
$\epsilon_{\alpha\beta} c_{1, \alpha} c_{2,\beta}$, with charge
$2e$. Tunnelling from a normal metal electrode or tip is predicted
to show a hard gap, despite ballistic, dissipationless in-plane
resistance. Conversely, tunnelling from a superconducting tip
should show zero gap.

A purely transport measurement is also possible using shot noise,
which has previously been used to probe fractional charges in
quantum Hall edge states~\cite{noise1,noise2,noise3,noise4}. By
introducing a quantum point contact, either using electrostatic
gates or a nano-constriction, edge-to-edge backscattering is
possible at that contact, with a finite transmission
probability~\cite{noise4}. Individual tunneling events will carry
charge $\pm 2e$, which is directly observable in the noise
spectrum. The detailed calculation about the shot noise in a
quantum point contact geometry has been presented in a follow-up
paper by some of the current authors~\cite{zhang2016}.

Here we propose a different method to measure the carrier charge
at the boundary. Compared with the point-contact geometry, our
current proposal is easier to implement experimentally, and more
convenient to analyze theoretically, as it only involves one edge
instead of two opposite edges. Our proposal is based on the
dual-gated geometry that has been used in experiments
Ref.~\onlinecite{young2013}. The screened Coulomb interaction in
our system can be tuned by its distance $d$ to the gates due to
screening. The competition between interaction and the Zeeman
energy can lead to a rich phase diagram, and when the interaction
is dominant, the system develops a canted antiferromagnetic (CAF)
order~\cite{young2013}. The size of the fermion gap at the
boundary, as well as the magnetic field required to realize the
BSPT state in this set up will be discussed in detail in the
supplementary material.

The stability of the edge states of our system relies on the
conservation of $S^z$, and if locally the $S^z$ conservation is
broken, the edge modes encounters backscattering, and hence leads
to noise of the current. We propose to screen the Coulomb
interaction for most of the sample, while leaving a region close
to the edge unscreened, in order to develop a local CAF order,
which serves as a local ``magnetic impurity" that breaks the $S^z$
conservation.
We calculate the quantum shot noise in the supplementary material
with the proposed set-up Fig.~\ref{structure}, and recover the
expected result: \beqn \tilde{S}(\omega=0)=2e^*\langle
I \rangle\coth \frac{e^*V}{2k_BT}. \eeqn 
$e^\ast = 2e$ is the smoking gun signature of the BSPT state
proposed in our work.




If a direct second order quantum phase transition between the BSPT
and trivial phase found in Ref.~\onlinecite{kevinQSH,mengQSH2}
indeed happens in a real system, then at the transition, which
corresponds to a $(2+1)d$ CFT, the bulk conductivity should be a
universal value $\sigma = D e^2/h$, where $D$ is an order-1
universal constant~\cite{fisher1,fisher2}. Moreover the transition
should be accompanied by a closing of the spin gap, with
observable consequences for spin susceptibility as well as thermal
transport measurements.

Zhen Bi, Yi-Zhuang You and Cenke Xu are supported by the David and
Lucile Packard Foundation and NSF Grant No. DMR-1151208; Leon
Balents is supported by NSF Grant No. DMR-1506119. Chao-Xing Liu
acknowledges the support from Office of Naval Research (Grant No.
N00014-15-1-2675), and helpful discussions with Jun Zhu. The first
two authors made equal contributions to this work.

\bibliography{graphene}

\begin{center}
{\bf Supplementary Materials}
\end{center}

\section{Experimental Conditions to realize and observe the Bosonic SPT state}

Based on the set-up we proposed in the main text, here we attempt
to estimate the dependence upon the distance to the inner dual
gates, $d$, of two key quantities: (1) the critical Zeeman
magnetic field $B_c$, required to make the {\em bulk} transition
from the CAF to FM state and (2) the fermion gap $\Delta_f$ at the
edge. Both are very difficult quantities to determine
quantitatively from pure theory, but we will make necessary
assumptions and use experimental input.

First consider the bulk phase transition. We assume the high field
limit so that the system is entirely within the zero energy Landau
levels (LLL) -- here the $n=0$ and $n=1$ levels due to the bilayer
nature. Then the Hamiltonian is the sum of the Zeeman energy and
the LLL projection of the interaction term. It is well-known that
the long-range part of the Coulomb interaction has SU(4) symmetry,
and hence does not distinguish the CAF and FM states. The energy
difference between them therefore comes from the anisotropic
contributions, which are short-range, i.e. originate at the
lattice scale. They can be represented in the continuum limit by a
two-body potential with a two-dimension delta-function dependence
with prefactor of order $e^2 a_0$, where $a_0$ is the lattice
constant. If we had {\em only} these anisotropic interactions and
no long-range Coulomb term, then the anisotropic interactions
projected into the LLL would yield interactions of order $V_{ani}
\sim c e^2 a_0/l_B^2$ per electron, where $c$ is an order 1
constant~\cite{caftheory}. $l_B$ is the magnetic length $l_B \sim
25.7\mathrm{nm} /\sqrt{B_z [\mathrm{Tesla}]}$, where $B_z$ is the
out-of-plane magnetic field.

A complication is that these short-range terms can be amplified by
the effects of the long-range interactions. In a field theoretic
perspective, the $1/r$ Coulomb interaction is marginal for
relativistic Dirac electrons, leading to renormalizations of
short-range (finite) interactions on scales where the relativistic
approximation applies. This has been discussed theoretically in
several contexts~\cite{Maxim,Gonz}. This effect {\em enhances} the
anisotropic short-range interactions. For bilayer graphene, this
is some range of energy scales between the inter-layer hopping
scale and some fraction of order one of the bandwidth. At scales
between the inter-layer hopping energy and the cyclotron energy,
there is a quadratic dispersion, and the problem is even more
complex, but further renormalization certainly occurs. Ultimately
the renormalizations are cut-off, in the absence of external
screening, by the magnetic length $l_B$ (or in energy, the
cyclotron energy).

Here we consider the effects of symmetric gates, placed a distance
$d$ above and below the bilayer. Charges in the bilayer
experienced a screened interaction due to an infinite array of
alternating image charges created by the gates. In the limit of
large separation of those charges, the screening becomes
exponential. One can show that two electric charges $e$ sandwiched
symmetrically between metal plates is given by
\begin{equation}
  \label{eq:1}
V(r) \sim \frac{2 e^2}{\sqrt{d r}}\exp(- \frac{\pi}{2}
\frac{r}{d}),
\end{equation}
when $r \gg d$.

This screening effect can cut off the renormalization of the
anisotropic interactions, so the renormalization is cut off by the
{\em smaller} of $l_B$ and the screening length of order $d$.  The
amount of renormalization is determined by the ratio of this
cut-off length to the lattice scale $a_0$. Very roughly, then, the
renormalized anisotropic interaction is of order
\begin{eqnarray}
  \label{eq:3}
  V_{ani}^R &  \sim &  c \frac{e^2 a_0}{\epsilon l_B^2} \left( \frac{{\rm Min}
  (l_B,A d)}{a_0}\right)^\eta \nonumber \\
  & \sim &  c \frac{e^2 a_0}{\epsilon l_B^2} \left( \frac{A^2 d^2 l_B^2}{a_0^2(A^2d^2+l_B^2)}\right)^{\eta/2}.
\end{eqnarray}
where $\epsilon$ is the dielectric constant, and $\eta$ is a
positive exponent proportional to the bare effective
fine-structure constant, which we will view as an unknown
parameter. In the second line we replace the ${\rm min}$ function
with a smoother interpolation.

The critical total magnetic field $B_T^c$ (including both inplane
field $B_{in}$ and out-of-plane field $B_z$) is determined just by
the balance of the anisotropy energy and the Zeeman energy. Hence,
$B_T^c \propto V_{ani}^R$ in Eq.~(\ref{eq:3}). Making the {\it ad
hoc} assumptions $\eta=1$ and $A=1$, we then have the functional
form of the critical field versus $d$,
\begin{equation}
  \label{eq:4}
  B_T^c = C\frac{e^2}{\epsilon l_B}\left( \frac{ d^2 }{d^2+l_B^2}\right)^{1/2},
\end{equation}
where $C$ is a constant independent of $d$ and $l_B$ which can be
fit to data, in for instance Ref.~\onlinecite{young2013}, where
the gates are far enough compared with the magnetic length, hence
the screening from the gates is weak.

\begin{figure}[t]
\centering
\includegraphics[width=210pt]{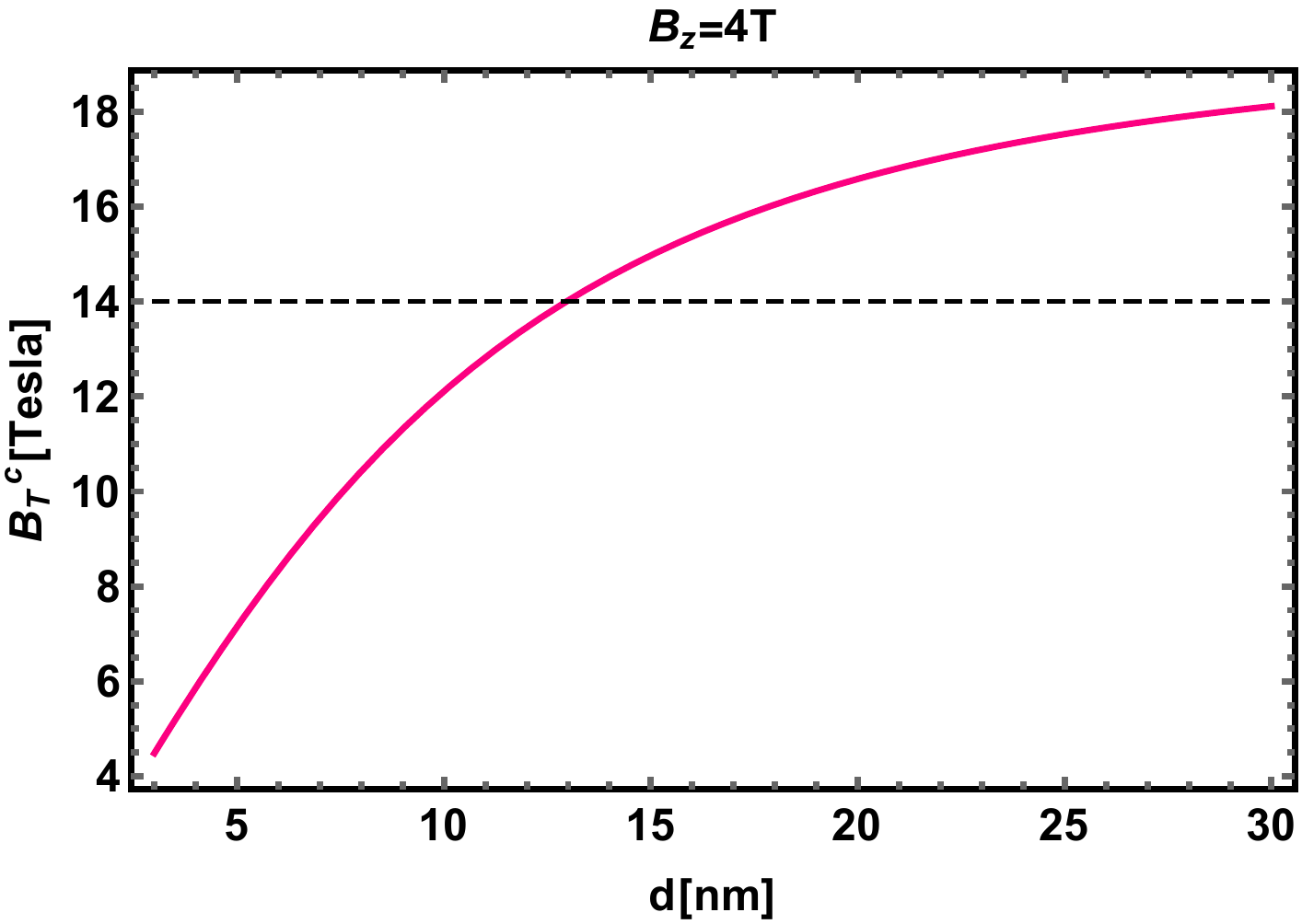}
\caption{The estimated critical $B_T^c$ at given $d$, with $B_z =
4$T. The constant $C$ in Eq.~\ref{eq:4} is fixed by experiment
Ref.~\onlinecite{young2013}, where the screening from the gates is
weak. So if the distance $d$ from the sample to the screening
gates is 8 nm, a 14T total field will produce the BSPT
(ferromagnetic state) in the screened region in Fig.2 of the main
text.}
  \label{plot1}
\end{figure}

\begin{figure}[t]
\centering
\includegraphics[width=210pt]{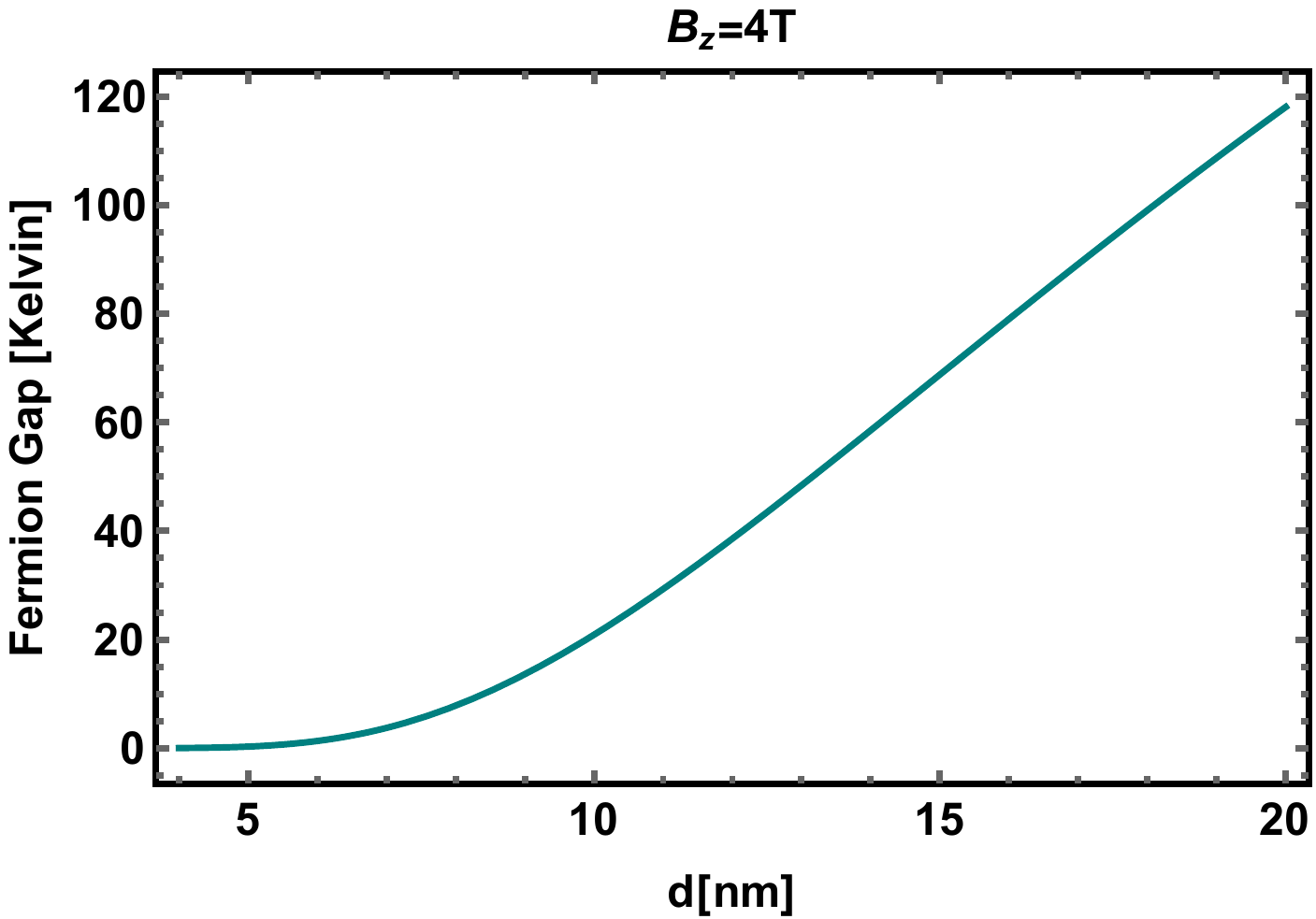}
\caption{The estimated fermion gap at the boundary at given $d$,
with $B_z = 4$T. We have taken $\epsilon = 11$~\cite{young2016}
and the constant $\tilde{C} = 1$ in Eq.~\ref{K}.}
  \label{plot2}
\end{figure}

Now we turn to the edge gap.  This is more difficult and less
universal because it depends upon the actual structure of the
edge, which depends upon the nature of the boundary, the potential
there, etc.  For example, if it is a physical edge of a graphene
sheet, the result will be different than for an edge defined
electrostatically. What one has to do in principle is to write the
screened Coulomb interaction, taking into any renormalizations due
to the aforementioned Coulombic effects at energies larger than
the cyclotron energy, and then project it into the manifold of
single-particle states defining the gapless edge modes near the
Fermi energy.  From this we obtain the interactions at the edge,
which include a forward-scattering contribution and the
inter-channel backscattering term (anharmonic term in Eq.6) of the
main text.  Then we should use the theory of the one-dimensional
edge including both these terms to estimate the gap.

The main difficulty is with the first step.  We do not expect the
Coulombic ``high energy'' renormalizations to be critical at the
edge, and will neglect that to a first approximation.  The SU(4)
symmetry of the system is anyway broken by the edge so that the
dominant long-range part of the Coulomb interaction will
presumably contribute there to both the forward-scattering and the
anharmonic terms of the 1d theory.  The forward-scattering terms
$U_{\rm intra}$ and $U_{\rm inter}$ in Eq.(4) of the main text
reflect long-distance interactions of the smooth part of the
electron density in the two channels.  Consequently, we expect
them to be given just by the zero momentum Fourier transform of
the screened Coulomb interaction.  We write
\begin{equation}
  \label{eq:5}
  U \sim \int \! dx\, V(x),
\end{equation}
and use the interpolating form
\begin{equation}
  \label{eq:6}
  V(x) =  \frac{e^2}{\epsilon \sqrt{l_B^2 + x^2}}
  \left(\frac{d^2+16x^2}{d^2}\right)^{1/4} e^{-\pi x/(2d)},
\end{equation}
which matches Eq.~(\ref{eq:1}) when $r \gg d$, shows pure Coulomb
behavior for $l_B<x<d$, and is regularized by the magnetic length
at short distances.  This gives in the two limits (with similar
forms for both $U$'s)
\begin{equation}
  \label{eq:2}
  U \sim \left\{ \begin{array}{lc} 2e^2/\epsilon \ln (d/l_B) & \textrm{for }
  d\gg l_B \\
  \\
  2.02 e^2/\epsilon \times d/l_B & \textrm{for } d \ll b_B \end{array}\right.
\end{equation}
So a reasonable interpolation is
\begin{equation}
  \label{eq:7}
  U \approx 2\frac{e^2}{\epsilon} \frac{d}{d+l_B},
\end{equation}
up to the weak logarithmic factor.

The anharmonic term $\alpha$ in Eq. (5) of the main text reflects
short-distance physics which allows scattering between channels.
It is much more sensitive to the structure of the edge state
wavefunctions.  We can proceed purely dimensionally by noting the
$\alpha$ is dimensions of energy divided by length.  Assuming that
it is primarily determined by distances of order $l_B$ itself, we
see that $\alpha \sim A e^2/(\epsilon l_B^2)$, with an unknown
prefactor $A$ which depends on all the edge details, if we neglect
screening. Screening will reduce this if $d<l_B$.  The form of
this screening is not really clear, and again how it affects the
value of $\alpha$ must depend upon edge details.  So we propose a
simple ad-hoc form for $\alpha$:
\begin{equation}
  \label{eq:8}
\alpha = A \frac{e^2}{\epsilon l_B^2} \left(
\frac{d}{d+l_B}\right).
\end{equation}

Now we are in a position to estimate the edge fermion gap.  The
scaling dimension of the anharmonic term is determined by the
Luttinger parameter as \beqn \Delta_{[\alpha]} = 2 -
\frac{2\pi}{K_{eff}}, \eeqn where $K_{eff}$ is the effective
Luttinger parameter {\em including} the forward-scattering
interactions.  Using the interpolation for $U$ in
Eq.~(\ref{eq:7}), we then estimated \beqn K_{eff} =
\sqrt{\left(\pi + \tilde{C} \frac{U}{\hbar v}\right)\pi},
\label{K}\eeqn where $\tilde{C}$ is another order-1 constant. $v$
is the velocity of the edge states, $\hbar v/l_B$ can be taken as
the cyclotron energy $\hbar \omega_c$ of bilayer graphene: $ \hbar
\omega_c \sim 3.2B_z[T]$ meV, which is also the ultraviolet energy
cut-off of the $1d$ edge system.

Taking the bare value of $\alpha$ in Eq.~(\ref{eq:8}), for small
$d$ we use standard methods to determine the fermion gap
$\Delta_f$ in the corresponding sine-Gordon model: \beqn \Delta_f
\sim \frac{\hbar v}{l_B} \left( \frac{\alpha}{\hbar v}
\right)^{1/\Delta_{[\alpha]}}. \eeqn This scaling form is given by
both the exact solution of the mass gap of the Sine-Gordon model,
and also the standard renormalization group flowing of the
anharmonic term.  At large $d$ we simply find a gap of order
$\hbar v/l_B$ or of order the cyclotron energy.  It is probably
somewhat smaller, but this reduction is hard to estimate, and
comes from the unknown prefactors, e.g. $A$ in Eq.~(\ref{eq:8}).


If we fix $B_{z}$ and increase $d$, since the interaction energy
scale will increase monotonically with $d$, so will the fermion
gap at the boundary.

For $B_z$ fixed at 4 Tesla, at given $d$, the estimate of critical
total field $B_T^c$, and also the corresponding fermion gap at the
boundary, are plotted in Fig.~\ref{plot1} and Fig.~\ref{plot2}. So
if the screening gates are made at 8 nm from the sample, a 14T
total field will produce the BSPT (ferromagnetic order) in the
screened region. In plot Fig.~\ref{plot2}, we have taken $\epsilon
= 11$~\cite{young2016} and $\tilde{C} = 1$ in Eq.~\ref{K}.

We also need another pair of outer gates to tune the strength of
the local CAF, to ensure the local CAF is not too strong to
completely block the transmission current. The strength of the CAF
can also be tuned by the distance $d'$ to the outer gates. The
renormalized anisotropic interaction inside the CAF domain can be
estimated using the same equation Eq.~\ref{eq:3}, where now $d$ is
replaced by the distance to the outer gates $d'$. For example for
$B_z$ fixed at 4 Tesla, and $d = 8nm$, we need $d'$ greater than
$13$ nm to create the CAF in the region not screened by the inner
gates.

\section{Single particle backscattering and instanton tunnelling}

Without interaction, our system reduces to two channels of quantum
spin Hall (QSH) insulator. A local CAF order at the edge of this
system will induce single particle backscattering. At the edge of
the system, the most general form of single particle
backscattering (SPB) is \bea
T_{l_1,l_2}&=&t_{l_1,l_2}\psi^{\dagger}_{l_1,L}\psi_{l_2,R}+h.c.
\cr\cr &\sim&
t_{l_1,l_2}\cos[\pi(\phi_{l_1}+\phi_{l_2})+(\theta_{l_1}-\theta_{l_2})],
\eea where $l_{1,2}\in \{1,2\}$ labels the two channels of the
fermion modes. As an example, we take $l_1=l_2=l$, and \bea
T_{l,l}&=&t_{l,l}\cos2\pi(\phi \pm \phi_-). \eea The sign of $\pm$
depends on whether $l$ takes $l=1$ or $l=2$. In the strong
scattering limit (or when the backscattering is relevant), we find
that \bea
&\phi=&\frac{(n+m+1)}{2}, \nonumber \\
&\phi_-=&\frac{(n-m)}{2}, \label{Eq:m-n def} \eea $i.e.$ $\phi$
and $\phi_-$ are pinned to discrete values with
$m,n\in\mathbb{Z}$. Quantum tunnelling generally happens between
different minima of the cosine potential, which is accompanied by
a non-zero change $\Delta n$ or $\Delta m$. This is known as the
instanton tunnelling events. The instanton tunnellings of $\phi$
and $\phi_-$ also corresponds to ``charge" currents associated
with fields $\phi$ and $\phi_-$ across the local CAF region. In
particular, $\phi$ carries the physical electric charge and thus
the minimal instanton tunnelling of $\phi$ characterizes the
current of the elementary electric charge of the system. In the
above SPB process, the minimal instanton tunnelling event of
$\phi$ field is given by $\Delta \phi=\frac{1}{2}$ or equivalently
$\Delta m+\Delta n=1$, which corresponds to the physical electric
current \bea
I_{\phi}=I_{\phi_1}+I_{\phi_2}=-e\partial_t(\phi_1+\phi_2)=-2e\partial_t
\phi \eea Therefore, the electric charge transport during this
minimal instanton event is given by \bea \Delta Q=\int dt
I_{\phi}=-2e\int dt \partial_t \phi=-2e\times\frac{1}{2}=-e. \eea
This result reveals that the elementary charge of fermionic QSH
system is $e$ electric charge.


\section{Minimal backscattering on a BSPT edge}

When the Coulomb interaction is turned on, the boundary of the
system becomes purely bosonic, as we showed in the main text. In
mathematical terms, the Coulomb interaction pins $\phi_{-}$ to a
fixed value (Here we choose $\phi=0$ or equivalently $n=m$). As a
result, the backscattering takes the following form: \bea T = t
\cos2\pi\phi. \eea In the strong scattering limit, $\phi$ is
pinned to $\frac{(2n+1)}{2}$ and the minimal instanton tunnelling
event is \bea \Delta \phi=1. \eea The elementary electric charge
transport now is thus given by \bea \Delta Q=-2e\Delta\phi=-2e.
\eea The $2e$-charge instanton event is a direct consequence of
the bosonic nature of the BSPT edge physics.

\begin{figure}[t]
\centering
\includegraphics[width=210pt]{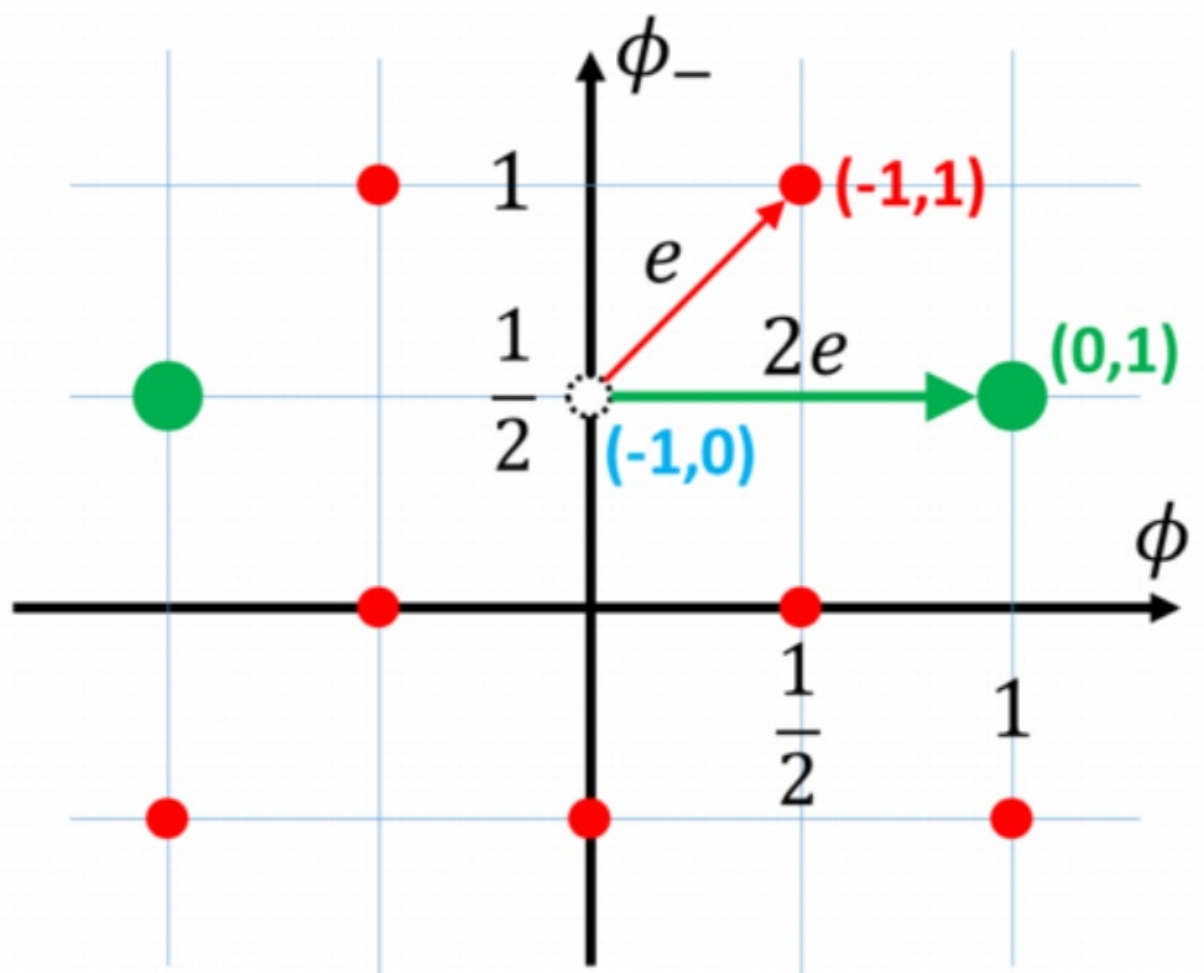}
\caption{Minimal instanton tunnelling events are depicted by (a) a
red arrow (charge-$e$ tunneling) for the QSH system and (b) a
green arrow (charge-$2e$ tunneling) for the BSPT system. Starting
from the dashed circle, red (green) dots are the possible
destinations of instanton tunnelling events for the noninteracting
quantum spin Hall (BSPT) system. }
  \label{Fig:Instanton tunneling}
\end{figure}

The instanton tunnelling charge can be visualized in Fig.
\ref{Fig:Instanton tunneling}, where we plot the bosonic field
configuration space of $\phi$ and $\phi_-$. We start with an
initial configuration shown by a dashed circle with
$(\phi,\phi_-)=(0,\frac{1}{2})$, which is equivalent to
$(m,n)=(-1,0)$. In the noninteracting bilayer quantum spin Hall
insulator, instanton tunnelling can happen between the dashed
circle and any colored dots (both red ones and green ones), and a
minimal instanton tunnelling with $(\Delta m,\Delta n)=(0,1)$ is
labelled by the red arrow, characterizing an $e$-charge tunnelling
process. In the BSPT limit, however, $\phi_-$ is pinned and does
not enter the instanton tunnelling process. This results in an
extra constraint, \bea \Delta m=\Delta n. \eea As a consequence,
the previous minimal instanton tunnelling is prohibited and  the
new minimal instanton tunnelling $(\Delta m, \Delta n)=(1,1)$ is
depicted by the green arrow in Fig. \ref{Fig:Instanton tunneling},
which corresponds to a $2e$-charge tunneling event.

In realistic systems, the rate of the instanton tunnelling event
$\tilde{t}$, also known as the fugacity of the instanton gas, is
rather small. As a result, any instanton event that involves
multiple instanton tunnellings will be suppressed. Therefore, we
expect that the minimal instanton tunnelling which induces $2e$
charge transport will be dominant in experimental detection.

\section{Current noise spectrum}

The charge of minimal instanton tunnelling can be directly read
out by measuring the noise spectrum of tunneling current. As shown
in Ref. [\onlinecite{martin2005,maciejko2009,zhang2016}], the
non-equilibrium current noise spectrum at zero frequency
$\tilde{S}(\omega=0)$ is defined as the anti-commutator of
tunneling current operator $I_{\eta}$, \bea
\tilde{S}(\omega=0)=\int d(t-t')\sum_{\eta}\langle {\cal T}_K
\{I_{\eta}(t),I_{-\eta}(t')\}\rangle + {\cal O}(\tilde{t_{l,l}}^2
), \eea where we have defined the Keldysh contour as $K$ and the
index $\eta=\pm$ to characterize the forward ($+$) and backward
($-$) branch. In the BSPT limit, the instanton tunnelling current
is given by \bea I_{\eta}(t)=4e\tilde{t}_{l,l}\sin
[2\tilde{\phi}(t)-2eVt], \eea where we have performed the duality
transformation and introduced a new dual field $\tilde{\phi}$
\cite{kane1992}. Here $V$ is the applied Voltage bias and
$\tilde{t}_{l,l}$ is the fugacity of the instanton gas. The
non-equilibrium current is, \bea \langle I(t)
\rangle=\frac{1}{2}\sum_{\eta}\langle {\cal T}_K I_{\eta}
e^{-2i\int_K dt_1 \tilde{t}_{l,l}\cos(2\tilde{\phi}_{+}-2eVt)}
\rangle. \eea By evaluating the Keldysh contour integral exactly,
we find that the relation between current noise spectrum
$\tilde{S}(\omega=0)$ and electric current $I$ due to instanton
tunnelling is given by \bea \tilde{S}(\omega=0)=2e^*\langle I
\rangle\coth \frac{e^*V}{2k_BT}, \eea with $e^*=2e$. In the
low-temperature/high-bias limit ($e^*V>>k_B T$), the Schottky
relation of quantum shot noise is recovered, \bea
\tilde{S}(\omega=0)=2e^*\langle I \rangle. \eea


\end{document}